\documentstyle[aps,prl,amsfonts,epsfig]{revtex}
\tightenlines
\def\be{\begin{equation}}
\def\ee{\end{equation}}

\title{A Continuous Transition Between
Quantum and Classical Mechanics (II)} \author{Partha Ghose and Manoj
K. Samal} \address{S. N. Bose National Centre for Basic Sciences,
Block JD, Sector III, Salt Lake, Kolkata 700 098, India}

\begin{document}
\maketitle
\begin{abstract}
Examples are worked out using a new equation proposed in the
previous paper to show that it has new physical predictions for
mesoscopic systems.
\end{abstract}
\vskip 0.2in


\section{Introduction}

If one assumes that quantum theory is more fundamental than
classical theory, then the emergence of a classical world from a
quantum substratum becomes a necessary requirement that quantum
theory must satisfy. It is not sufficient to show the emergence of
a macroscopic world that is only approximately or partially
classical in nature. The accuracy of the prediction of energy loss
from binary pulsars by the general theory of relativity is just as
impressive as that of quantum electrodynamics for the Lamb
shift\cite{will}. Unfortunately, nobody has yet succeeded in
demonstrating satisfactorily that quantum theory passes this test.
The approach that comes closest to doing justice to the problem is
that of environment-induced decoherence (EID) \cite{eid}.
Unfortunately, it too falls short of doing full justice because it
leaves the notorious measurement problem unsolved. In this paper
we will show that a different approach combined with the idea of
decoherence is able to bridge the gap between quantum and
classical theory smoothly and exactly \cite{ghose1} and leads to
new predictions.

The most striking feature of quantum theory that distinguishes it
from classical theory is the appearance of coherence exhibited by
interference phenomena. The idea of EID is based on the
observation that the environment surrounding a quantum system can,
in effect, monitor some of the system's observables in such a
fashion that the eigenstates of those observables continuously
decohere, i.e. the environment not only causes states with
macroscopic separations to decohere but also prevents such
coherence from reappearing. In technical terms, this is reflected
in the fact that the reduced density matrix of the system
(obtained by tracing over the environment variables) becomes
diagonal in the chosen representation over a time scale that is
set by the system parameters, the rate of relaxation and the
temperature of the environment (treated as a heat bath). However,
this falls short of ensuring that the surviving diagonal elements
of the density matrix behave classically. The usual arguments
based on properties of the Wigner distribution are deceptive.
Although it is true that the time evolution of the Wigner function
looks identical to that of the classical Liouville function for
all polynomial potentials of order no higher than the second,
nevertheless the energy spectra, for example, clearly continue to
exhibit the same quantum features as before EID. We will give a
number of concrete examples of this and other differences below.

This shortcoming of the EID approach is due to the fact that it is
based on the linear and unitary Schr{\"o}dinger evolution. It is,
however, well known that a unitary linear evolution can never lead
to a measurement in quantum theory which requires an independent
nonlinear, non-unitary process (von Neumann's "process 1"). This
is the crux of the measurement problem that remains unsolved. And
unless this is solved or obviated in some way, the gap between
quantum theory and classical theory cannot be bridged.
Consequently, we would like to explore in detail a different
approach proposed in the preceding paper.

The central point of this theory is based on the simple observation
that it is possible to give a unified description of classical and
quantum theory by introducing a wave function for a classical ensemble
of systems. Such a system is described by an action function
$S(x,t)_{cl}$ that satisfies the classical Hamilton-Jacobi equation

\begin{equation}
\frac{\partial S(x,t)_{cl}}{\partial t} + \frac{1}{2m} ( \nabla
S(x,t)_{cl})^2 + V(x) = 0
\label{eq:A}
\end{equation}
and an equation of continuity

\begin{equation}
\frac{\partial R^2 (x,t)_{cl}}{\partial t} + {\rm div} (R^2_{cl} (x,t)
\frac{p}{m}) = 0
\label{eq:B}
\end{equation}
for the position distribution function $R^2 (x,t)_{cl}$, where $p =
\nabla S(x,t)_{cl}$ is the momentum. Let us introduce a
wave function

\begin{equation}
\psi(x,t)_{cl} = R(x,t)_{cl} \: e^{ \frac{i}{\hbar} S(x,t)_{cl}}
\end{equation}
For example, consider an ensemble of free particles in one dimension and let

\begin{eqnarray}
R(x,t)_{cl} &=& (2 \pi \sigma^2 )^{-1/4} e^{- (x - vt)^2 /4
\sigma^2}\\ S(x,t)_{cl} &=& p x - E t \label{fgod}
\end{eqnarray}
This describes an ensemble in which the particles have the same
velocity $p/m$ and a position distribution that is Gaussian with a
spread $\sigma$ that does not change with time. In other words,
$\psi(x,t)_{cl}$ is a non-dispersive wave function that behaves
classically.

Another physically interesting example is the harmonic
oscillator. Consider the wave packet for which

\begin{eqnarray}
R(x,t)_{cl}&=& (\frac{m \omega}{\pi \hbar})^{\frac{1}{4}}
e^{- \frac{m \omega}{2 \hbar}(x - a \cos \omega t)^2}\\
S(x,t)_{cl}&=& \frac{1}{4}m \omega a^2
\sin 2 \omega t - m \omega x a \sin \omega t
\end{eqnarray}
The equation of motion of the particles (obtained from equation
(\ref{eq:A})) in the standard fashion) is

\begin{equation}
\frac{d p}{d t}= - \nabla V = - m \omega^2 x
\end{equation}
whose solution is $x = a \cos \omega t$. The particles in the
ensemble therefore behave exactly classically in spite of the fact
that they are described by a wave function that involves $\hbar$.

The equation of motion that these wave functions must satisfy so that
the classical equations (\ref{eq:A}) and (\ref{eq:B}) are preserved is
most interesting. It is

\begin{equation}
i\hbar \frac{\partial \psi(x,t)_{cl}}{\partial t}=
[-\frac{\hbar^2}{2m} \nabla^2 + V(x) - Q_{cl}] \psi(x,t)_{cl}
\label{eq:C}
\end{equation}
with

\begin{equation}
Q_{cl}= -\frac{\hbar^2}{2m} \frac{\nabla^2 R(x,t)_{cl}}{R(x,t)_{cl}}
\end{equation}
This is a nonlinear equation that reduces to the Schr\"{o}dinger
equation in the limit $Q_{cl}$ going to zero. $Q_{cl}$ acts like a
potential that destroys linearity and therefore coherence, and makes
the system behave exactly classically, {\it in spite of the occurrence
of Planck's constant in the equation}. This is a point that is worth
emphasizing. The traditional approach has been an attempt to show that
quantum mechanics reduces to classical mechanics in the limit $h$
going to zero. This has not been entirely successful. However, the
fact is that $h$ {\it is not zero in the real world}. So, the real
problem is to explain how the macroscopic world behaves classically
{\it in spite of Plank's constant not being zero}. This has not been
nearly sufficiently emphasized in the literature.

We therefore take the point of view advocated in the preceding
paper that classical mechanics is a theory in which the Planck
constant $\hbar$ occurs as a fundamental constant through equation
(\ref{eq:C})\footnote{We do not wish to speculate on the origin of
fundamental constants. We simply postulate them, as in all
physical theories so far proposed.}. The systems nevertheless
behave completely classically because of the (hitherto ignored)
fundamental interaction $Q_{cl}$ which is nonlinear and cancels
all quantum features exactly. Note that $Q_{cl}$ remains `hidden'
in the usual formulation of classical mechanics because it is not
formulated in terms of a wave function. Its form is not {\it ad
hoc}. It is uniquely fixed by the equations (\ref{eq:A}) and
(\ref{eq:B})), i.e., by the sole requirement of classicality.
Quantum theory then emerges naturally and in full glory in the
limit that this new fundamental interaction is switched off. This,
we believe, is a more reasonable and natural assumption than the
usual one that an effective nonlinear term such as $Q_{cl}$ does
somehow arise from a basically linear equation such as the
Schr\"{o}dinger equation. We fully realize that ours will not be
an immediately popular view, and most of our readers familiar with
the reigning paradigm of physics will be skeptical. We only wish
to remind them that our point of view, though perhaps different,
is not inconsistent with that of Bohr, the founder of the
Copenhagen philosophy. Bohr always emphasized that the classical
nature of the measuring apparatus cannot be derived from quantum
mechanics. This is in sharp contrast to the prevalent view that it
is possible to understand the emergence of the classical world
starting from quantum theory. Our experience with the heretical
point of view we are pursuing has convinced us that its
consequences cannot be dismissed off hand, and need to be
seriously explored and confronted with experiments. We believe it
is crazy enough that it has a chance of being right.

The theory described so far is an almost unknown way of describing
classical physics. It is interesting but admittedly sterile in the
sense that it contains no new physical result. The new physics enters
through the postulate that $Q_{cl}$ is the effective potential seen by
a classical system that is surrounded by an environment. Equation
(\ref{eq:C}) should then be replaced by the more fundamental equation

\begin{equation}
i\hbar \frac{\partial \psi(x,t)}{\partial t}= [-\frac{\hbar^2}{2m}
\nabla^2 + V(x) - \lambda(t) Q] \psi(x,t) \label{eq:D}
\end{equation}
where the coupling parameter $0 < \lambda (t)\leq $ and
encapsulates the cumulative effect of random disturbances caused
by the environment on the system. This type of cumulative effect
of the environment on a system is generic and occurs in a wide
class of phenomena. The limit $\lambda(0) \rightarrow 0$
corresponds to quantum mechanics (pure Schr\"{o}dinger equation)
provided we also assume that $\psi$ is single-valued, and the
limit $\lambda(\infty)=1$ corresponds to classical physics
(equation (\ref{eq:C})). The two worlds are smoothly bridged by
this equation. However, its specific new predictions really come
in the intermediate ranges of $\lambda(t)$ corresponding to
mesoscopic systems, and are weakly dependent on the model of the
system-environment coupling. But before showing that, we will now
demonstrate how this equation makes the diagonal terms of the
reduced density matrix of a system behave exactly classically
after decoherence.

Before concluding this section, we would like to point out that
there are other approaches too to decoherence in which the
Schr\"odinger equation is modified by the addition of ad hoc
non-linear, non-unitary terms that cause `explicit collapse'. Our
approach is totally different in that the primacy of the
mesoscopic world as opposed to the quantum world makes it
unnecessary for us to invoke the concept of collapse. The
empirical differences between models of `explicit collapse' and
EID (`false collapse') have been discussed by Ghose\cite{ghose2}.

\section{EID: Conventional View}

Consider a wave function that is initially a superposition of two
different position eigenfunctions:

\begin{equation}
\psi(x,0)= \frac{1}{\sqrt{2}}[\psi_1(x,0) + \psi_2(x,0)]
\end{equation}
Its density matrix $\rho(x,x',0)$ has off-diagonal terms. In the
presence of an environment simulated by a heat bath at a
temperature $T$, the time evolution of the density matrix is
governed by a master equation of the form \cite{eid}

\begin{equation}
 \frac{d \rho}{d t} = - \frac{i}{\hbar} [H, \rho] - \gamma (x - x')
(\partial_x - \partial_{x'}) \rho - \rho/\tau_D
\label{eq:E}
\end{equation}
where $\gamma$ is the relaxation time and $\tau_D = h^2 /2 m
\gamma k_B T (x - x')^{2} = \gamma^{-1} (\lambda_T/ \Delta x)^2$ where
$\lambda_T$ is the thermal de Broglie wavelength. The first term
describes the usual quantum evolution, the second term causes
dissipation due to friction and the third term is responsible for
fluctuations or Brownian motion. It is the last term that is
principally responsible for decoherence of states, whose positions are
macroscopically distinguishable, over a time scale $\tau_D$ which
could be as small as $10^{-23}$ sec for a mass of $1$ gm at room
temperature ($T= 300$ K) even if the relaxation time $\tau_R =
\gamma^{-1}$ is of the order of $10^{17}$ sec, the age of the
universe. On the other hand, for microscopic particles like the
electron $\tau_D$ could be much larger than $\tau_R$ on atomic and
larger time scales. Notice that these last two terms responsible for
decoherence according to the prevalent view are both {\it
off-diagonal} (they vanish when $x = x' $) and have no effect on the
first (diagonal) term {\it which remains quantum mechanical}.

We will now give two examples of this. Consider a superposition of
the harmonic oscillator stationary state $n = 1$ centered around
$x = x_1$ and $x = x_2$,

\begin{equation}
\psi_{os} (x, t) = (\frac{m w}{\pi \hbar})^{1/4} [e^{- \frac{m
\omega (x - x_1)^2}{2 \hbar}} (x - x_1) + e^{- \frac{m \omega (x -
x_2)^2}{2 \hbar}} (x - x_2)],
\end{equation}
as well as that of a dispersive, free Gaussian quantum mechanical
wave packet,
\begin{eqnarray}
\psi_{fg}(x,t)& =& \frac{1}{\sqrt{2}} (2 \pi s_t^2)^{-1/4} [ e^{ [
i k (x- x_1 - u t/2) - (x - x_1 - u t)^2 /4 s_t \sigma_0]} \\
\nonumber & & + e^{ [ i k (x- x_2 - u t/2) - (x - x_2 - u t)^2 /4
s_t \sigma_0]} ]\label{fgq}
\end{eqnarray}
with $ s_t = \sigma_0 (1 + i\hbar t/ 2 m \sigma_0^2)$. It is clear
from these expressions that although the off-diagonal terms of the
corresponding density matrices die out because of decoherence,
{\it the diagonal terms remain intact}. Although the environment
will prevent the off-diagonal terms from reappearing over an
effectively infinite time scale, the energy spectra of the
remaining diagonal terms will still reveal their initial quantum
nature. For example, it is straightforward to see that the
harmonic oscillator energy will continue to be $3 \hbar/2$ for the
$n=1$ stationary state, and
\begin{eqnarray}
<E> = \frac{p^2}{2 m} + \frac{\hbar^2}{8 m \sigma_0^2}
\end{eqnarray}
for the free Gaussian wave packet. In the latter case it is also
clear that the diagonal terms will continue to be dispersive and
so maintain their quantum character.

Exactly the same conclusions regarding the energy spectra also
follow from the Wigner distributions
\begin{equation}
W(x,p) = \frac{1}{2 \pi \hbar} \int_{- \infty}^{\infty} e^{i p y/\hbar}
\rho (x - \frac{y}{2}, x + \frac{y}{2}) dy
\end{equation}
(for convenience defined here in one space dimension only) in
these cases. The fact that the time evolution of the Wigner
function happens to coincide with the classical Liouville
function,
\begin{eqnarray}
\frac{d W}{d t}&=& \frac{\partial W}{\partial t} + \frac{p}{m}
\frac{\partial W}{\partial x} - \frac{\partial V}{\partial
x}\frac{\partial W}{\partial p} \\ \nonumber & = & \frac{\partial
W}{\partial t} + \{ W, H\}_{PB}
\end{eqnarray}
for polynomial potentials of degree less than or equal to two, as
already mentioned, is therefore clearly misleading. In fact, one
can show that for stationary states of the harmonic oscillator the
energy spectrum is given in terms of the Wigner function by the
expression\cite{Carruthers}

\begin{equation}
E = \frac{p^2 }{2 m} + \frac{1}{2} m \omega^2 x^2 - \frac{\hbar^2}{8 m
W} \frac{\partial^2 W}{\partial x^2} - \frac{ m \omega^2 \hbar^2}{8 W}
\frac{\partial^2 W}{\partial p^2}
\end{equation}
The third and fourth terms are clearly non-classical. It is
straightforward to verify this for $n = 1$ and $n = 2$. In these
cases one again obtains the expected quantum mechanical results,
$E = 3 \hbar \omega/2, 5 \hbar \omega/2$ for $n = 1$ and $n = 2$
respectively.

Before concluding this section, let us just note the expression
for the rate of change of $W(x, p)$ that follows from the master
equation (\ref{eq:E}),
\begin{equation}
\frac{d W}{d t} = - \{W, H \}_{PB} + 2 \gamma \frac{\partial(p W)}
{\partial p} + D \frac{\partial^2 W}{\partial p^2}\label{eq:z}
\end{equation}
where $D = 2 m \gamma k_B T$. It is the last term in this equation
that causes the off-diagonal terms to disappear by a process of
diffusion.

\section{EID: Heretical View}

A consequence of the new equation (\ref{eq:D}) is that the density
matrix $\rho$ must now satisfy a different evolution equation,
namely

\begin{eqnarray}
\frac{d \rho}{d t} &=& - \frac{i}{\hbar} [H', \rho] \\
H'& =& -\frac{\hbar^2}{2m} \nabla^2 + V(x) - \lambda (t) Q
\label{eq:F}
\end{eqnarray}
This clearly shows that the potential $\lambda (t) Q$ will modify
the density matrix and ensure its exact classical behaviour in the
limit $\lambda = 1$. To see this in terms of the Wigner function,
it will be convenient to impose the de Broglie-Bohm guidance
condition $p = m d x/d t = \nabla S$. Then one has
\begin{eqnarray}
E &=& - \frac{\partial S}{\partial t} = \frac{p^2}{2 m} + V(x) +
(1 - \lambda (t)) Q\\ \frac{d p}{d t} &=&  - \nabla [V(x) + (1 -
\lambda (t)) Q]
\end{eqnarray}
The rate of change of the Wigner function can then be written in
the form

\begin{eqnarray}
\frac{d W}{d t} &=& \frac{\partial W}{\partial t} + \frac{p}{m}
\frac{\partial W}{\partial x} - \frac{\partial W}{\partial p}
\frac{\partial E}{\partial x} + ...\nonumber\\ &=& \frac{\partial
W}{\partial t} + \{ W, E \}_{PB} + ...\nonumber\\ &=&
\frac{\partial W}{\partial t} + \{ W, E_{cl} \}_{PB} + (1 -
\lambda (t)) \{W, Q \}_{PB} + ... \label{w}
\end{eqnarray}
where the dots represent the conventional EID terms. When $\lambda
\rightarrow 0$, this includes the effect of the quantum potential
$Q$ and is therefore non-classical. However, in the limit $\lambda
= 1$ this reduces to the familiar classical equation in terms of
the Poisson bracket. Since there is complete equivalence between
de Broglie-Bohm theory and standard quantum mechanics for Gibbs
ensembles, this completes the demonstration that our theory causes
complete decoherence by reducing the Wigner function to its
genuine classical form. The de Broglie-Bohm theory turns out to be
the most convenient and natural framework for establishing this
connection between classical and quantum theory.

It is clear from equations (\ref{eq:z}) and (\ref{w}) that the
conventional theory of EID and our theory lead to different physical
predictions. It should therefore be possible, in principle, to
test which of these theories corresponds more closely to nature.

\section{Bohmian and Classical Trajectories}

In this section we will give three examples of how Bohmian
trajectories smoothly pass over to classical trajectories on complete
decoherence. Let us first note that the general equation of motion for
the trajectories that follows from equation (\ref{eq:D}) is

\begin{equation}
m \frac{d^2 x}{d t^2} = - \nabla [V(x) + (1 - \lambda (t)) Q]
\label{eq:G}
\end{equation}
It is very difficult to solve this equation analytically because
of the nonlinearity introduced by the potential $Q$. We will
therefore solve it numerically, and for this purpose will make use
of an approximation that satisfies the correct boundary
conditions. We will replace $Q$ in this equation by the pure `
quantum potential', i.e., by the expression for $Q$ in the quantum
mechanical limit $\lambda = 0$ which is easy to calculate for a
number of physically interesting cases. Then, equation
(\ref{eq:G}) will still reduce to the classical equation in the
limit $\lambda = 1$, and to the de Broglie-Bohm equation in the
quantum limit $\lambda = 0$.

Before we can actually compute the trajectories, we need a model
for the system-environment coupling. We shall assume that the
environment acts in a completely random manner on the system so
that it 'decays' exponentially to the classical system
($\lambda = 1$). The, a convenient choice for the parameter
$\lambda (t)$ is $\lambda (t) = 1 - \exp (- b t)$ with
$b = 1/\tau_D$.

Let us now consider stationary states of the harmonic oscillator
for which the action function is $S = - E_n t$ with $E_n = (n +
1/2) \hbar \omega$. Then, the momentum $p = \partial S/\partial x
= 0$, and the particles in the ensemble are all at rest. (This
does not, of course, mean that their measured momenta will be
zero. For details, see Holland (\cite{Holland}, Chapter 8.) Figure
1(i) shows the trajectories for the states $n =0, 1$ and $2$. The
quantum potential in this case is easily calculated from the
quantum Hamilton-Jacobi equation to be

\begin{equation}
Q = (n + \frac{1}{2}) \hbar \omega - \frac{1}{2} m \omega^2 x^2
\end{equation}
Equation (\ref{eq:G}) can then be solved numerically for
appropriate values of the parameters $m, \omega$ and $b$. (We have
used the units $\hbar=c=1$ and taken $m=1$ MeV, $\omega=1$, $t$ in
units of $10^{21}$ sec and therefore $b$ in units of $10^{-21}$
sec$^{-1}$). The trajectories are shown in Figures 1(ii), 1(iii)
and 1(iv). It is gratifying to see how the trajectories for the
stationary states smoothly go over to the oscillating modes in the
classical limit (large $b$).

\begin{figure}[htbp]
\centerline{\epsfbox{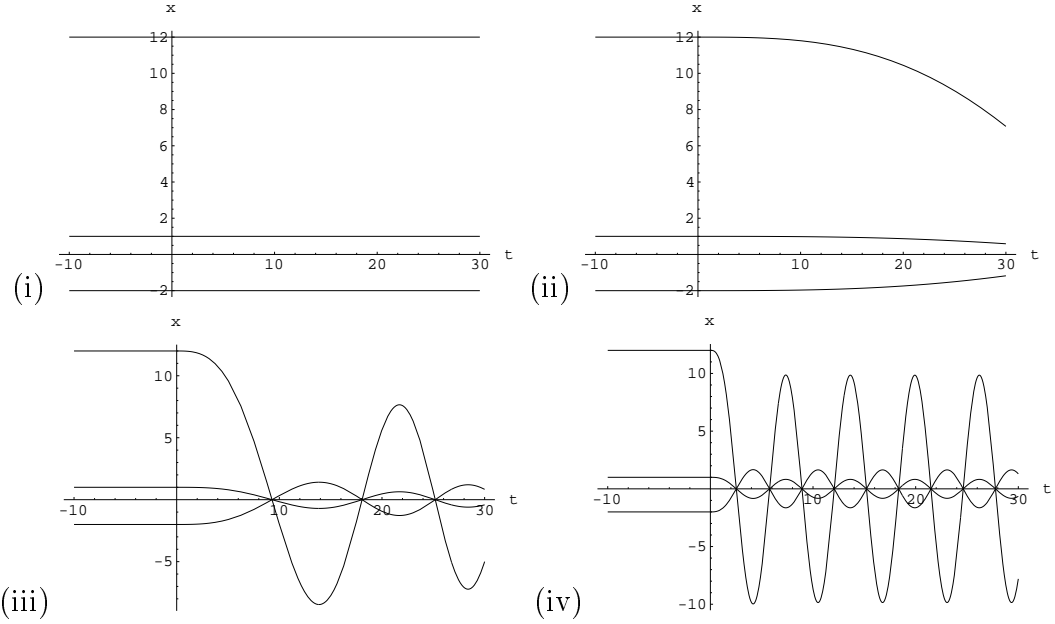}} \caption{x vs. t for the
quantum oscillator (stationary state) for (i) b=0, (ii) b=0.0001,
(iii) b= 0.01 and (iv) b=0.7 }
\end{figure}

Let us next consider the non-dispersive Gaussian wave packet
solution of the quantum harmonic oscillator:
\begin{eqnarray}
R(x,t)&=& (\frac{m \omega}{\pi \hbar})^{\frac{1}{4}} e^{[- \frac{m
\omega}{2 \hbar}(x - a \cos \omega t)^2]} \\ \nonumber S(x,t)&=& -
\frac{1}{2} \hbar \omega t + \frac{1}{4}m \omega a^2 \sin 2 \omega
t - m \omega x a \sin \omega t
\end{eqnarray}

Its quantum potential is given by \cite{Holland}

\begin{equation}
Q = - \frac{1}{2} m \omega^2 (x - a \cos \omega t)^2 + \frac{1}{2}
\hbar \omega
\end {equation}
Equation (\ref{eq:G}) can be numerically integrated in this case
also (in the same units as before with $\sigma_0 \approx 10^{-10}$
MeV$^{-1}$) The quantum orbits (corresponding to $\lambda = 0$)
are shown in Figure 2(i). Over the ensemble (obtained by taking
different initial positions) the particles oscillate over
different centre points and cross the classical amplitude $a$.
Note that their trajectories do not cross. The trajectories after
decoherence are shown in Figures 2(ii), 2(iii), and 2(iv). They
remain within the classical amplitude $a$ but cross one another,
as expected.
\begin{figure}[htbp]
\centerline{\epsfbox{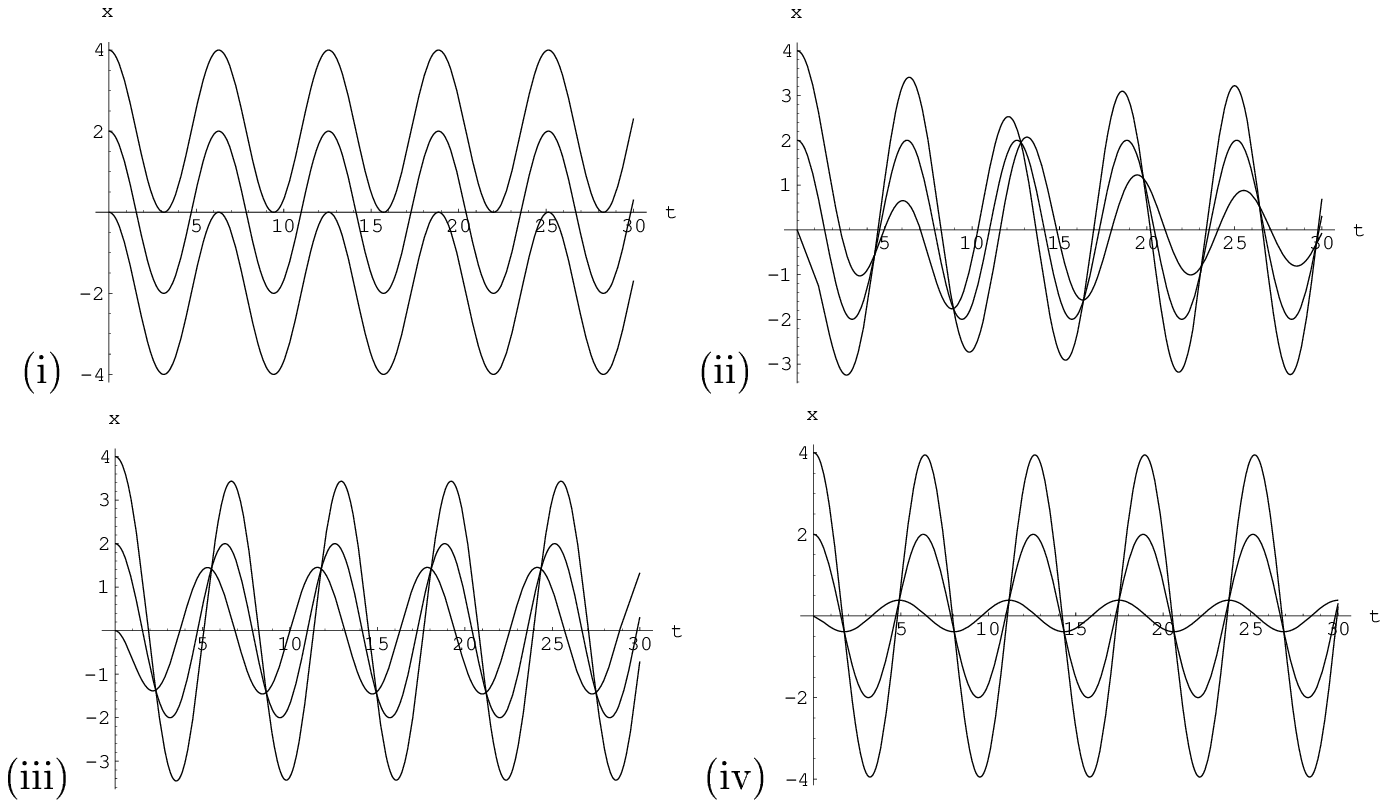}} \caption{x vs. t for the
quantum oscillator (non-stationary state) for (i) b=0, (ii) b=0.1,
(iii) b= 1 and (iv) b=5 }
\end{figure}

Finally, let us consider the one dimensional free wave packet
corresponding to eqn. (\ref{fgod}). Then,

\begin{eqnarray}
R(x,t) &=& (2 \pi \sigma^2 )^{-1/4} e^{- (x - ut)^2 /4 \sigma^2}\\
S(x,t) &=& -\frac{\hbar}{2} \tan^{-1} (\frac{\hbar t}{2 m
\sigma_0^2}) + m u (x - \frac{1}{2} u t) + (x - u t)^2
\frac{\hbar^2 t^2}{8 m \sigma_0^2 \sigma^2}
\end{eqnarray}
with $ \sigma^2 = \sigma_0^2 [1 + \frac{\hbar^2 t^2 }{4 m^2
\sigma_0^4}] $. Such a wave packet spreads in time. Its quantum
potential is given by

\begin{equation}
Q = \frac{\hbar^2}{4 m \sigma^2} [1 - (x - u t)^2 /2 \sigma^2 ]
\end{equation}
Again it is possible to solve equation (\ref{eq:G}) numerically in
this case. The quantum mechanical trajectories for a packet at
rest ($u = 0$) are shown in Figure 3(i). They are hyperbolic
trajectories that do not cross. The packet contracts for $t<0$ and
expands for $t>0$. The 'meso' and classical trajectories are
shown in Figures 3(ii), 3(iii), and 3(iv).

\begin{figure}[htbp]
\centerline{\epsfbox{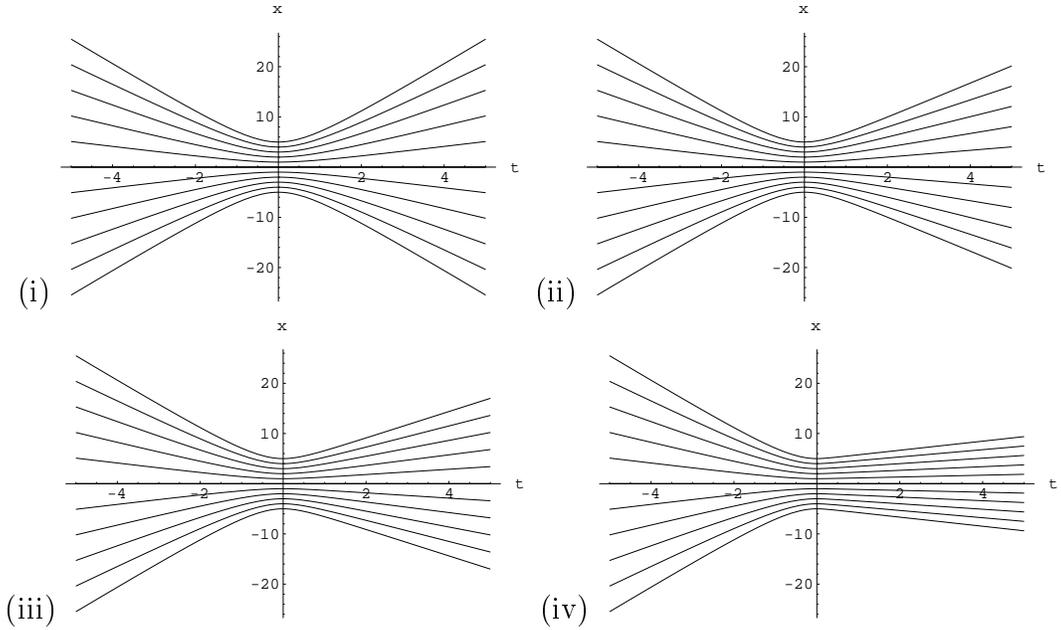}}
\caption{x vs. t for the free
gaussian wavepacket for (i) b=0, (ii) b=0.5, (iii) b= 1 and (iv)
b=5 }
\end{figure}

Figure 4 is a plot of the velocities as functions of time. It is
clear from these plots that the accelerations cease and the
velocities become constant in the $t > 0$ region : the
trajectories become parallel and the packet does not spread any
more.

\begin{figure}[htbp]
\centerline{\epsfbox{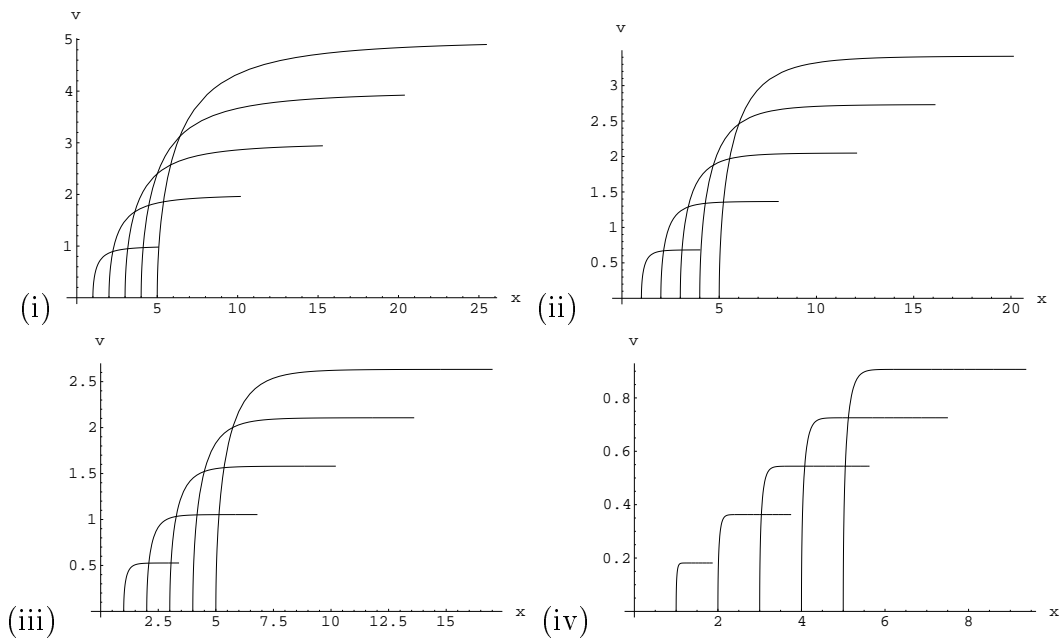}} \caption{v vs. x for the free
gaussian wavepacket for (i) b=0, (ii) b=0.5, (iii) b= 1 and (iv)
b=5 }
\end{figure}

There is clearly an asymmetry in all these cases between $t < 0$
and $t > 0$, as expected. All this offers a totally new insight
into the connection between Bohmian and classical trajectories,
beautifully exhibited in these figures.
\section{Conclusions}

The smooth transition between quantum (i.e., Bohmian) and
classical trajectories exhibited in the previous section are new
results that have interesting physical consequences. The
mesoscopic systems implied by the theory proposed in the previous
paper and elaborated in this paper are a new prediction in the
sense that they are conceptually different from "mesoscopic"
states described by quantum mechanics. According to our theory
these new systems are neither fully quantum nor fully classical
but provide a smooth link between the two. If the new formulation
is correct, it should be possible to observe, for example,
oscillations of tiny cantilevers and study how these oscillations
change as their coupling to the environment is varied, and compare
them with the predicted curves (like those in Fig. 2). It should
be possible to construct such cantilevers\cite{Zala}. Other
physical mesoscopic systems are under study at present using the
new formulation.

If this formulation turns out to be correct, it would imply that
the de Broglie-Bohm theory is, in a sense, more fundamental than
standard quantum theory. Although these two theories are
physically completely equivalent when Gibbs ensembles are used,
nevertheless the physical interpretations are radically different.
Furthermore, the conventional de Broglie-Bohm theory can be
non-ergodic in certain multi-particle cases whereas the
corresponding standard quantum mechanical systems are ergodic
\cite{ghose3}. This implies that the time average of at least one
observable in such systems is different from its space average in
the de Broglie-Bohm theory, leading to a direct conflict with
standard quantum theory in which the two averages are the same. A
realistic two-photon experiment is under preparation at Torino to
distinguish between standard quantum mechanics and the de
Broglie-Bohm theory. The same experiment would also give a clear
verdict on the truth or falsity of the new formulation.

\section{Acknowledgement}
The authors thank the Department of Science \& Technology, Govt.
of India for financial support to carry out this work.

\end{document}